\newcommand{\be}{\begin{equation}}
\newcommand{\ee}{\end{equation}}
\newcommand{\bea}{\begin{eqnarray}}
\newcommand{\eea}{\end{eqnarray}}

\newcommand{\vecvar}[1]{\mbox{\boldmath$#1$}}
\newcommand{\ri}{{\rm i}}
\newcommand{\rd}{{\rm d}}
\newcommand{\re}{{\rm e}}
\newcommand{\ket}[1]{\left|#1\right\rangle}
\newcommand{\bra}[1]{\left\langle#1\right|}
\newcommand{\braket}[2]{\left\langle#1|#2\right\rangle}
\newcommand{\Braket}[2]{\left\langle#1\Biggm|#2\right\rangle}
\newcommand{\eref}[1]{eq.~(\ref{#1})}
\newcommand{\Eref}[1]{Equation (\ref{#1})}
\newcommand{\fref}[1]{Fig.~\ref{#1}}
\newcommand{\Fref}[1]{Figure \ref{#1}}

\documentclass[letter]{jpsj2} 
\title{Temporal Oscillation of Conductances in 
Quantum Hall Effect of Bloch Electrons}

\author{
Manabu \textsc{Machida}$^{1}$
\thanks{Email address: machida@iis.u-tokyo.ac.jp}, 
Naomichi \textsc{Hatano}$^{1}$
\thanks{E-mail address: hatano@iis.u-tokyo.ac.jp} and 
Jun \textsc{Goryo}$^{2}$
\thanks{E-mail address: jungoryo@phys.aoyama.ac.jp}
}

\inst{$^{1}$Institute of Industrial Science, The University of Tokyo, 
Komaba, Meguro, Tokyo 153-8505 \\
$^{2}$Department of Physics and Mathematics, Aoyama Gakuin University, 
5-10-1 Fuchinobe, Sagamihara, Kanagawa 229-8558}

\abst{
We study a nonadiabatic effect on the conductances in the 
quantum Hall effect of two-dimensional electrons with a periodic 
potential.  
We found that the Hall and longitudinal conductances oscillate 
in time with very large frequencies due to quantum fluctuation.  
}

\kword{integer quantum Hall effect, Chern number, 
nonadiabatic effect, Greenwood linear-response theory}

\begin{document}
\maketitle


Two years after the discovery of the quantum Hall effect
\cite{Kawaji80a,vonKlitzing80a}, 
Thouless, Kohmoto, Nightingale and den Nijs (TKNN)
\cite{Thouless82a,Kohmoto85a} theoretically studied 
the quantum Hall effect in two-dimensional electrons with a periodic 
potential.  
They expressed the Hall conductance as an integer multiplied 
by $e^2/h$.  
The integer is a topological invariant called the Chern number.  
Such a system has been experimentally realized as a superlattice 
structure in a semi-conductor heterojunction.
\cite{Liu91a,Geisler04a,Geisler05a}  

We can treat the electric field as a time-dependent vector potential.  
Then adiabatic approximation gives the same expression of 
the Hall conductance as in the TKNN theory.\cite{Kohmoto89a,Hatsugai97a}  
In this Letter, we consider the conductances of the same system taking 
a nonadiabatic effect into account.  
We determine the quantum fluctuation of the Hall conductance $\sigma_{xy}$ and 
the longitudinal conductance $\sigma_{yy}$; \textit{i.e.}, 
$\sigma_{xy}$ and $\sigma_{yy}$ oscillate in time with very large 
frequencies.  

We consider the nonadiabatic effect on the conductances 
with the help of the Greenwood linear-response theory\cite{Greenwood58a} 
and study a correction to the adiabatic approximation.  
Wagner studied the failure of the Kubo formula due to a nonadiabatic 
effect on the dc current in a finite system.
\cite{Wagner92a,Wagner92b}  
Grimaldi \textit{et al.} explained the high-$T_{\rm c}$ superconductivity 
of fullerene compounds by taking nonadiabatic effects into 
account.\cite{Grimaldi95a}  

This Letter is organized as follows.  We first calculate the 
conductances beyond the adiabatic approximation and obtain 
correction terms to $\sigma_{xy}$ and $\sigma_{yy}$ of the TKNN 
theory.  We then evaluate the oscillations in 
these correction terms numerically.


We consider noninteracting electrons in a periodic potential in 
the $x$-$y$ plane.  
A magnetic field is applied in the $z$ direction, and 
an electric field is applied in the $y$ direction.  
We treat the electric field as a vector potential.  
Using the Landau gauge, we express the Hamiltonian of the system as 
\be
\mathcal{H}(t)=\frac{1}{2m_{\rm e}}\left(\vecvar{p}+e\vecvar{A}(t)\right)^2
+U(\vecvar{x}),
\ee
where
\bea
\vecvar{A}(t) &=& \left(Bx-E_yt\right)\vecvar{e}_y,
\nonumber \\
U(\vecvar{x})&=& 
U_x\cos\left(\frac{2\pi x}{a}\right)+U_y\cos\left(\frac{2\pi y}{b}\right).
\eea

We consider the flux per unit cell of $U(\vecvar{x})$ to be the rational 
number $p/q$ in the unit of the flux quantum, which produces $p$ 
subbands.  We label each subband by $m\;(1\le m\le p)$ hereafter.  
Then we obtain the generalized crystal momentum $\hbar\vecvar{k}$ defined 
in the magnetic Brillouin zone.\cite{Goryo02c} 
$0\le k_x< 2\pi/qa$ and $0\le k_y< 2\pi/b$.  

We define $\mathcal{H}_{\vecvar{k}}(t)$ as 
\be
\mathcal{H}_{\vecvar{k}}(t) \equiv 
\re^{-\ri\vecvar{k}\cdot\vecvar{x}}\mathcal{H}(t) 
\re^{\ri\vecvar{k}\cdot\vecvar{x}}.
\ee
The instantaneous Hamiltonian $\mathcal{H}_{\vecvar{k}}(t)$ 
is diagonalized as 
\be
\mathcal{H}_{\vecvar{k}}(t)\ket{u_{m\vecvar{k}}(\vecvar{x},t)} =
\epsilon_{m\vecvar{k}}(t)\ket{u_{m\vecvar{k}}(\vecvar{x},t)}.
\ee
Since 
$\mathcal{H}_{\vecvar{k}}(t)=\mathcal{H}_{k_x,k_y-eE_yt/\hbar}(0)$, 
we obtain for $m\neq n$
\bea
\epsilon_{m\vecvar{k}}(t) &=& \epsilon_{m\vecvar{k}}(0)+O(E_y), 
\nonumber \\
\braket{u_{m\vecvar{k}}(\vecvar{x},t)}{
\dot{u}_{n\vecvar{k}}(\vecvar{x},t)}
&=& \nonumber \\
&& \hspace{-40mm}\frac{eE_y}{\hbar}
\Braket{\frac{\partial u_{m\vecvar{k}}(\vecvar{x},0)}{\partial k_y}}
{u_{n\vecvar{k}}(\vecvar{x},0)}
+O\left(E_y^2\right),
\eea
where the dot indicates the time derivative.  
We note that, by fixing the arbitrary phase in each time, we can always 
achieve $\braket{u_{m\vecvar{k}}(\vecvar{x},t)}{
\dot{u}_{m\vecvar{k}}(\vecvar{x},t)}
=0.$

The time evolution of the density operator is given by the 
von Neumann equation
\be
\frac{\rd}{\rd t}\rho(t)=\frac{1}{\ri\hbar}[\mathcal{H}(t),\rho(t)].
\ee
Following the Greenwood linear-response theory,\cite{Greenwood58a} 
we expand the density operator with respect to the electric field and 
take the zeroth- and first-order terms into account 
\be
\rho(t)=\rho^{(0)}+E_y\rho^{(1)}(t).
\ee
Thus, the matrix elements 
\be
\rho_{mn\vecvar{k}}\equiv
\bra{u_{m\vecvar{k}}(\vecvar{x},t)}
\re^{-\ri\vecvar{k}\cdot\vecvar{x}}\rho(t)\re^{\ri\vecvar{k}\cdot\vecvar{x}}
\ket{u_{n\vecvar{k}}(\vecvar{x},t)}
\ee
are approximated using 
\bea
\rho_{mn\vecvar{k}} &=&
\rho_{mn\vecvar{k}}^{(0)}+E_y\rho_{mn\vecvar{k}}^{(1)} \nonumber \\
&=& f_{m}\delta_{mn}+E_y\rho_{mn\vecvar{k}}^{(1)},
\eea
where $f_{m}\;(=f(\epsilon_{m\vecvar{k}}))$ is the Fermi distribution.  
Hereafter, we let $\epsilon_{m\vecvar{k}}$ denote 
$\epsilon_{m\vecvar{k}}(0)$ and $\ket{u_{m\vecvar{k}}}$ denote 
$\ket{u_{m\vecvar{k}}(\vecvar{x},0)}$.  

The time evolution of the off-diagonal elements ($m\neq n$) is 
given by 
\be
\frac{\rd}{\rd t}\rho_{mn\vecvar{k}}^{(1)}
=
\frac{1}{\ri\hbar}(\epsilon_{m\vecvar{k}}-\epsilon_{n\vecvar{k}})
\rho_{mn\vecvar{k}}^{(1)}
+\frac{e}{\hbar}(f_{m}-f_{n})
\Braket{\frac{\partial u_{m\vecvar{k}}}{\partial k_y}}
{u_{n\vecvar{k}}}.
\label{rhoeom}
\ee
Therefore, we obtain the matrix elements of the density operator as 
\bea
\rho_{mn\vecvar{k}}^{(1)}
&=&
-\ri e
\Braket{\frac{\partial u_{m\vecvar{k}}}{\partial k_y}}
{u_{n\vecvar{k}}}
\frac{f_{m}-f_{n}}
{\epsilon_{m\vecvar{k}}-\epsilon_{n\vecvar{k}}}
\left[
1-\re^{-\ri(\epsilon_{m\vecvar{k}}-\epsilon_{n\vecvar{k}})t/\hbar}
\right] \quad (m\neq n),
\label{rhomato}\\
\rho_{mm\vecvar{k}}^{(1)} &=& \mbox{const.}
\label{rhomatd}
\eea

Let us calculate the current $J_{\alpha}\;(\alpha=x,y)$ using 
\bea
J_{\alpha} &\equiv& 
{\rm Re}\left[
\langle J_{\alpha}(E_y)\rangle-\langle J_{\alpha}(0)\rangle
\right], 
\label{Jval}\\
\langle J_{\alpha}(E_y)\rangle
&=&
{\rm Tr}\left(\rho(t)
\frac{\partial\mathcal{H}(t)}{\partial A_{\alpha}(t)}\right) \nonumber \\
&=&
\frac{e}{\hbar} {\rm Tr}\left(\rho_{\vecvar{k}}(t)
\frac{\partial\mathcal{H}_{\vecvar{k}}(t)}{\partial k_{\alpha}}\right).
\label{Jop}
\eea
By plugging eqs.~(\ref{rhomato}) and (\ref{rhomatd}) into 
\eref{Jop}, we obtain 
\be
J_{\alpha}=
\frac{e^2}{\hbar}E_y {\rm Re} \int_{\rm MBZ}\frac{\rd^2\vecvar{k}}{(2\pi)^2}
\sum_{m\neq n}\ri(f_m-f_n)
\Braket{\frac{\partial u_{m\vecvar{k}}}{\partial k_y}}{u_{n\vecvar{k}}}
\Braket{u_{n\vecvar{k}}}{\frac{\partial u_{m\vecvar{k}}}{\partial k_{\alpha}}}
\left[
1-\re^{-\ri(\epsilon_{m\vecvar{k}}-\epsilon_{n\vecvar{k}})t/\hbar}
\right], 
\label{current}
\ee
where MBZ denotes the magnetic Brillouin zone.  
After some tedious but straightforward calculation, we obtain 
the conductances $\sigma_{xy}$ and $\sigma_{yy}$ in the forms 
\bea
\sigma_{xy}
&\equiv&
\frac{J_x}{E_y} \nonumber \\ 
&=&
\frac{e^2}{2\pi\hbar}N_{\rm Ch}
+
\frac{e^2}{\hbar} \int_{\rm MBZ}\frac{\rd^2\vecvar{k}}{(2\pi)^2} 
\sum_{m\neq n}(f_{m}-f_{n})r_{mn}(\vecvar{k})
\sin{\left[(\epsilon_{m\vecvar{k}}-\epsilon_{n\vecvar{k}})t/\hbar-
\theta_{mn}(\vecvar{k})\right]},
\nonumber \\
\sigma_{yy} 
&\equiv&
\frac{J_y}{E_y} \nonumber \\ 
&=&
\frac{e^2}{\hbar} \int_{\rm MBZ}\frac{\rd^2\vecvar{k}}{(2\pi)^2} 
\sum_{m\neq n}\left|
\Braket{\frac{\partial u_{m\vecvar{k}}}{\partial k_y}}{u_{n\vecvar{k}}}
\right|^2
\left(f_m-f_n\right)
\sin{(\epsilon_{m\vecvar{k}}-\epsilon_{n\vecvar{k}})t/\hbar},
\nonumber\\
&&
\label{fullsigma}
\eea
where $r_{mn}(\vecvar{k})$ and $\theta_{mn}(\vecvar{k})$ are the real 
numbers that satisfy 
\be
r_{mn}(\vecvar{k})\re^{\ri\theta_{mn}(\vecvar{k})}\equiv
\Braket{\frac{\partial u_{m\vecvar{k}}}{\partial k_y}}{u_{n\vecvar{k}}}
\Braket{u_{n\vecvar{k}}}{\frac{\partial u_{m\vecvar{k}}}{\partial k_x}}.
\ee
Here, $N_{\rm Ch}$ is the Chern number with a finite-temperature 
correction 
\be
N_{\rm Ch}
\equiv 
\sum_m f_{m}\int_{\rm MBZ}\frac{\rd^2\vecvar{k}}{\pi} 
{\rm Im}\left[
\Braket{\frac{\partial u_{m\vecvar{k}}}{\partial k_y}}
{\frac{\partial u_{m\vecvar{k}}}{\partial k_x}}\right].
\ee
We note that if we ignore the time dependence of 
$\rho_{mn\vecvar{k}}^{(1)}$ in \eref{rhoeom}, we obtain 
$\sigma_{xy}=(e^2/2\pi\hbar)N_{\rm Ch}$ and $\sigma_{yy}=0$, which 
are consistent with the TKNN theory.  

The above-mentioned equations imply that we obtain sinusoidally 
oscillating terms 
in addition to the Chern-number term in the Hall conductance 
$\sigma_{xy}$ due to quantum fluctuation.  The longitudinal conductance 
$\sigma_{yy}$ also oscillates with zero mean.  
The oscillation period is determined by the energy-gap size.  

\Eref{fullsigma} is also a gauge invariant.  We observe this by treating 
the electric field as the scalar potential 
$A_0(t)=-eE_yy\exp[-\ri\omega t]$ instead of the vector potential.  
The first-order time-dependent perturbation theory gives 
eqs.~(\ref{current}) and (\ref{fullsigma}) in the limit of 
$\omega\to 0$.  


To demonstrate the oscillation numerically, we assume that 
the state lies in the lowest Landau level.  
Moreover, we assume that the flux $\Phi\;(=abB)$ is given by the unit 
flux $\Phi_0\;(=2\pi\hbar/e)$ multiplied by the rational number $p/q$ as 
\be
\frac{\Phi}{\Phi_0}=\frac{abeB}{2\pi\hbar}=\frac{p}{q},
\ee
where $p$ and $q$ are coprime.  Due to the periodic potential 
$U(x,y)$, each Landau level splits into $p$ subbands with a $q$-fold 
degeneracy in each subband.  

In the weak-potential limit $|U(\vecvar{x})|\ll\hbar eB/m_{\rm e}$, 
the perturbation theory gives the instantaneous eigenfunction 
$\ket{u_{m\vecvar{k}}(\vecvar{x})}$ as\cite{Thouless82a} 
\bea
\ket{u_{m\vecvar{k}}(\vecvar{x})}
&=&
\frac{1}{\sqrt{\mathcal{N}}}
\sum_{\nu=1}^{p}d_m^{\nu}(\vecvar{k})
\sum_{\ell=-\infty}^{\infty}
\exp\left[
-\left(\frac{eB}{2\hbar}\right)
\left(x+\frac{\hbar k_y}{eB}-\ell qa-\frac{\nu qa}{p}\right)^2
\right]
\nonumber \\
&\times& 
\exp\left[-\ri k_x\left(x-\ell qa-\frac{\nu qa}{p}\right)\right]
\re^{-2\pi\ri y\frac{\ell p+\nu}{b}},
\eea
where $\mathcal{N}$ is the normalization factor.  
Here, the coefficients $\{d_{m}^{\nu}\}$ and first-order energy 
shift $\epsilon^{(1)}_{m\vecvar{k}}$ are determined by the following 
secular equation. 
\be
\beta d_{m}^{\nu-1}(\vecvar{k})+\alpha_{\nu}d_{m}^{\nu}(\vecvar{k})+
\beta^*d_{m}^{\nu+1}(\vecvar{k})
=\epsilon^{(1)}_{m\vecvar{k}} d_{m}^{\nu}(\vecvar{k}).
\label{secular}
\ee
Here,
\bea
\alpha_{\nu} 
&=& 
U_x\re^{-\pi qb/2pa}\cos{(-qbk_y/p+2\pi\nu q/p)},
\nonumber \\
\beta
&=&
\frac{U_y}{2}\re^{-\pi qa/2pb}\re^{-\ri qak_x/p}.
\eea
Note that the unperturbed energy 
$\epsilon^{(0)}_{m\vecvar{k}}$ is independent of 
$m$ and $\vecvar{k}$; \textit{i.e.}, 
$\epsilon_{m\vecvar{k}}-\epsilon_{n\vecvar{k}}=
\epsilon^{(1)}_{m\vecvar{k}}-\epsilon^{(1)}_{n\vecvar{k}}$.  
After straightforward calculation, we obtain
\bea
\Braket{\frac{\partial u_{m\vecvar{k}}}{\partial k_x}}
{\frac{\partial u_{m\vecvar{k}}}{\partial k_y}}
&=&
\sum_{\nu=1}^p \left(\frac{\rd d_m^{\nu}}{\rd k_x}\right)^*
\frac{\rd d_m^{\nu}}{\rd k_y}-\ri\frac{\hbar}{eB}, 
\\
\Braket{u_{n\vecvar{k}}}{\frac{\partial u_{m\vecvar{k}}}{\partial k_x}}
&=&
\sum_{\nu=1}^p \left(d_n^{\nu}\right)^*\frac{\rd d_m^{\nu}}{\rd k_x},
\\
\Braket{u_{n\vecvar{k}}}{\frac{\partial u_{m\vecvar{k}}}{\partial k_y}}
&=&
\sum_{\nu=1}^p \left(d_n^{\nu}\right)^*\frac{\rd d_m^{\nu}}{\rd k_y}.
\eea
Thus, we can calculate the conductances $\sigma_{xy}$ and $\sigma_{yy}$ 
in \eref{fullsigma} using the eigenvalues and eigenvectors 
in \eref{secular}.  

As typical values, we use $a=100{\rm nm}$, $b=130{\rm nm}$, and 
$U_x=U_y=0.1{\rm meV}$.  We assume $p/q=65/2$, implying 
$B=10.4{\rm T}$.  
With this flux ratio, the lowest Landau level splits into 65 subbands.  
We note that the $p$ subbands in the lowest Landau level range from 
$0.240{\rm meV}$ to $0.948{\rm meV}$.  The center of the next Landau 
level is located at $1.78{\rm meV}$.  
In the numerical calculation, we obtain the derivatives of 
$d_m^{\nu}(\vecvar{k})$ by the finite-difference method.  We consider the 
slices $\Delta k_x$ and $\Delta k_y$ to be $2\pi/100qa$ and 
$2\pi/100b$, respectively.  
The Chern number term is obtained by 
Fukui, Hatsugai and Suzuki's method\cite{FHS}.  

\begin{figure}[tb]
\begin{center}
\includegraphics[scale=0.25]{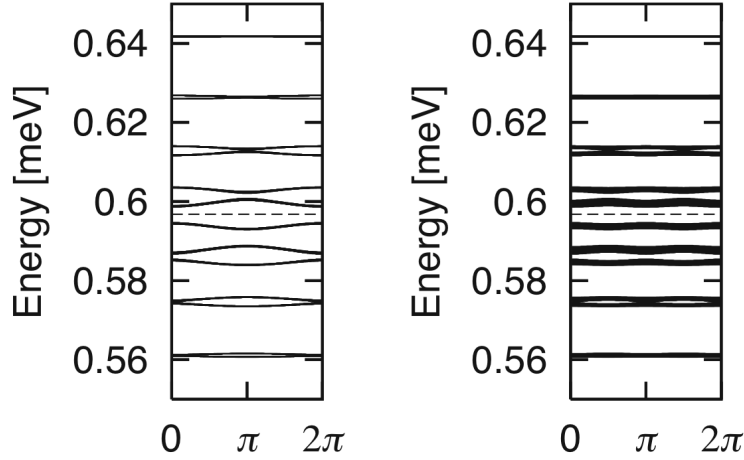}
\end{center}
\caption{The energy levels around the center of the $p\;(=65)$ subbands 
are shown as functions of (a) $qak_x$ and (b) $bk_y$.  The flux ratio 
$p/q=65/2$ corresponds to $B=10.4{\rm T}$.  We use 
$U_x=U_y=0.1{\rm meV}$, $a=100{\rm nm}$, and 
$b=130{\rm nm}$.  The Fermi energy (the dashed line) 
lies between the 33rd and 34th levels.}
\label{f33a}
\end{figure}

\begin{figure}[tb]
\begin{center}
\includegraphics[scale=1.8]{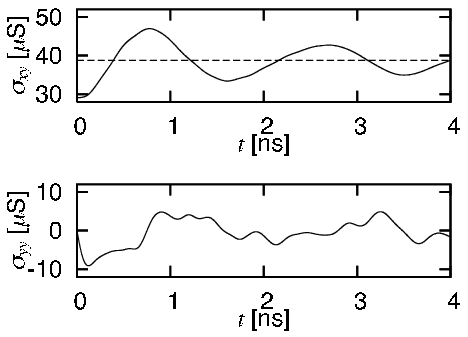}
\end{center}
\caption{Conductances corresponding to the situation in \fref{f33a}.  
In the upper panel, the solid line shows $\sigma_{xy}$ 
as a function of time, and the dashed line shows 
the Chern number term 
$\sigma_{\rm Ch}=(e^2/2\pi\hbar)N_{\rm Ch}$.  
In this case, $N_{\rm Ch}=1$.  
The lower panel shows $\sigma_{yy}$ as a function of time.  
}
\label{f33b}
\end{figure}

\begin{figure}[tb]
\begin{center}
\includegraphics[scale=0.25]{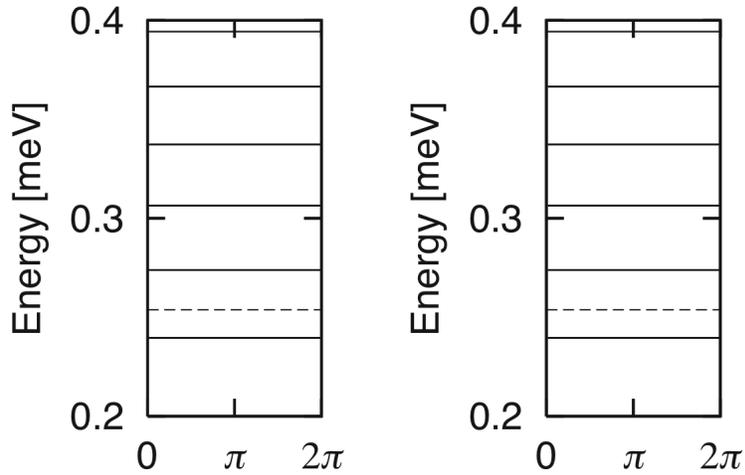}
\end{center}
\caption{The energy levels in the lowest part of the $p\;(=65)$ subbands are 
shown as functions of (a) $qak_x$ and (b) $bk_y$.  The flux ratio 
$p/q=65/2$ corresponds to $B=10.4{\rm T}$.  We use 
$U_x=U_y=0.1{\rm meV}$, $a=100{\rm nm}$, and 
$b=130{\rm nm}$.  The Fermi energy (the dashed line) 
lies between the second and third levels.  
In each panel, the lowest solid line below the dashed line expresses 
the degenerate first and second levels.}
\label{f02a}
\end{figure}

\begin{figure}[tb]
\begin{center}
\includegraphics[scale=1.8]{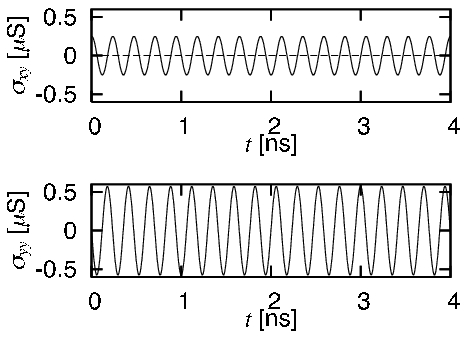}
\end{center}
\caption{
Conductances corresponding to the situation in \fref{f02a}.  
In the upper panel, the solid line shows $\sigma_{xy}$ 
as a function of time, and the dashed line shows 
the Chern number term 
$\sigma_{\rm Ch}=(e^2/2\pi\hbar)N_{\rm Ch}$.  
In this case, $N_{\rm Ch}=0$.  
The lower panel shows $\sigma_{yy}$ as a function of time.  
}
\label{f02b}
\end{figure}

First, we make the Fermi energy $\epsilon_{\rm F}$ lie between the 
33rd and 34th levels, as shown in \fref{f33a}.  
The fluctuations of $\sigma_{xy}$ and $\sigma_{yy}$ in \fref{f33b} 
consist of the oscillations corresponding to 
the energy gaps between levels higher than $\epsilon_{\rm F}$ and 
levels lower than $\epsilon_{\rm F}$.  
\Fref{f33b} shows that the time resolution required to detect the 
quantum fluctuation is about $10 {\rm ps}$.  
Secondly, we make $\epsilon_{\rm F}$ lie between the second and third 
levels, as shown in \fref{f02a}.  
Then we obtain almost periodic oscillations of $\sigma_{xy}$ and 
$\sigma_{yy}$ (\fref{f02b}) in contrast to the seemingly random 
oscillation in the former case.  This is because the energy gaps are 
almost constant in the latter case.  

To summarize, we studied a nonadiabatic effect on the conductances in the 
TKNN theory.  We found that both $\sigma_{xy}$ and $\sigma_{yy}$ 
oscillate due to quantum fluctuation.  The oscillation period 
is of the order of $100{\rm ps}$ to $1{\rm ns}$.  
It is a challenging but interesting issue to detect 
these oscillations experimentally.  

Although we studied the time dependence of currents at a 
constant electric field, by similar calculation, we observe that 
constant currents in reverse produce a time-dependent electric field.  
In this case, the oscillation period is also determined by 
the energy-gap size.  

\section*{Acknowledgments}
The authors thank Professor Tomoki Machida for valuable comments 
from an experimental point of view.  
One of the authors (M. M.) thanks Takahiro Fukui and Mikito Koshino 
for indispensable discussions on the numerical calculation of 
the Chern numbers.  
M. M. also thanks Seiji Miyashita for 
fruitful discussions on the Greenwood linear-response theory.  
The study is partially supported by a Grant-in-Aid for Scientific 
Research (No.~17340115) from the Ministry of Education, Culture, 
Sports, Science and Technology 
as well as by Core Research for Evolutional Science and Technology 
of Japan Science and Technology Agency.

\end{document}